\documentclass[twocolumn,prd,nofootinbib,superscriptaddress,amsmath]{revtex4}
\usepackage{graphicx}
\usepackage{bm}

\def\ms{\ensuremath{\overline{\rm MS}}}
\def\vev#1{\left\langle T\Bigl\{ \, #1 \Bigr\} \right\rangle}
\def\nn{\nonumber \\ }
\def\jir{\ensuremath{G_{1\gamma{\rm I}}}}
\def\jtot{\ensuremath{G_{\rm tot}}}
\newcommand{\3}[1]{\mathbf{#1}}  
\newcommand\slashed[1]{#1\llap{/}}

\begin{document}

\title{Renormalization of the Vector Current in QED}

\author{John C. Collins}
\affiliation{Physics Department, Penn State University, University
  Park, PA 16802}

\author{Aneesh V. Manohar}
\affiliation{Department of Physics, University of California at San Diego,
9500 Gilman Drive, La Jolla, CA 92093-0319}

\author{Mark B. Wise}
\affiliation{California Institute of Technology, Pasadena, CA 91125}

\date{May 9, 2006}

\begin{abstract}
  It is commonly asserted that the electromagnetic current is
  conserved and therefore is not renormalized.  Within QED we show (a)
  that this statement is false, (b) how to obtain the renormalization
  of the current to all orders of perturbation theory, and (c) how to
  correctly define an electron number operator.  The current mixes with
  the four-divergence of the electromagnetic field-strength tensor.  The
  true electron number operator is the integral of the time component
  of the electron number density, but only when the current differs
  from the \ms-renormalized current by a definite finite
  renormalization.  This happens in such a way that Gauss's law holds:
  the charge operator is the surface integral of the electric field at
  infinity.  The theorem extends naturally to any
  gauge theory.
\end{abstract}

\maketitle

\section{Introduction}

A classic statement about the electromagnetic current in quantum field
theory is that because it is conserved it needs no renormalization ---
see, e.g., \cite[p.\ 341]{sterman}, \cite[p.\ 430]{peskin.schroeder},
\cite[Sec.\ 10.1]{pokorski}, and \cite[p.\ 16]{hqbook}.  In fact this
result is false, as we will explain in detail.  Under
renormalization, the current may mix with operators of equal or lower
dimension whose four-divergence vanishes identically, i.e., without use
of the equations of motion.  As observed in \cite[p.\ 162]{collins} there are 
no such operators in the absence of gauge fields, and so the
nonrenormalization theorem holds in nongauge theories.  The theorem
also extends to pure QCD for the flavor currents, where the available
operators all have nonzero color.

\begin{figure*}
\centering
\begin{tabular}[t]{c@{\hspace*{15mm}}c@{\hspace*{15mm}}c}
   \includegraphics[scale=0.25]{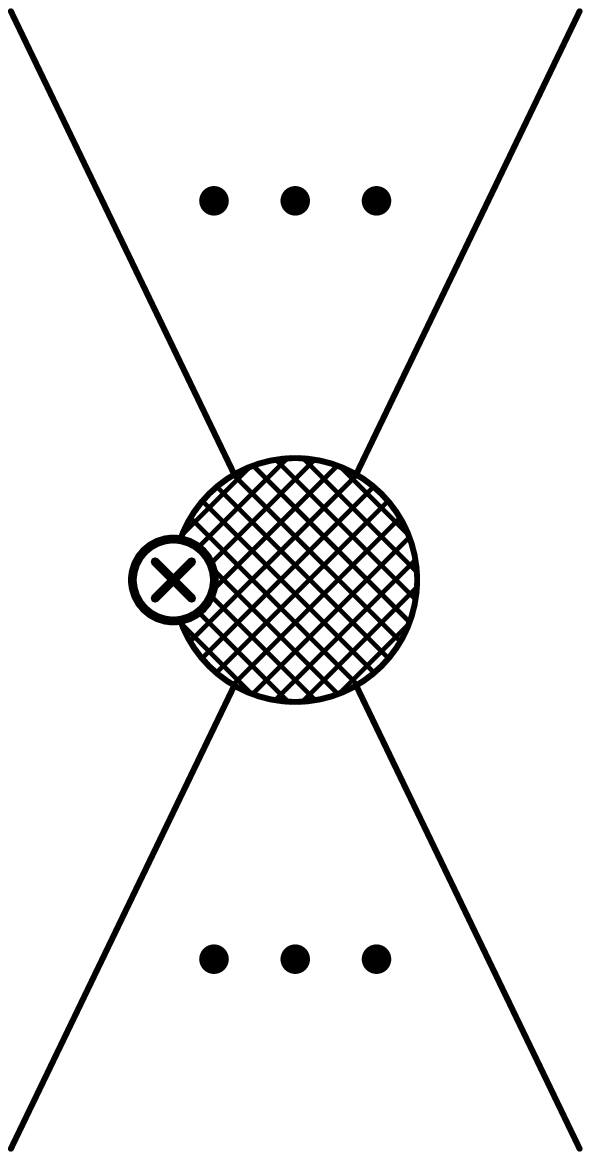}
   &\includegraphics[scale=0.25]{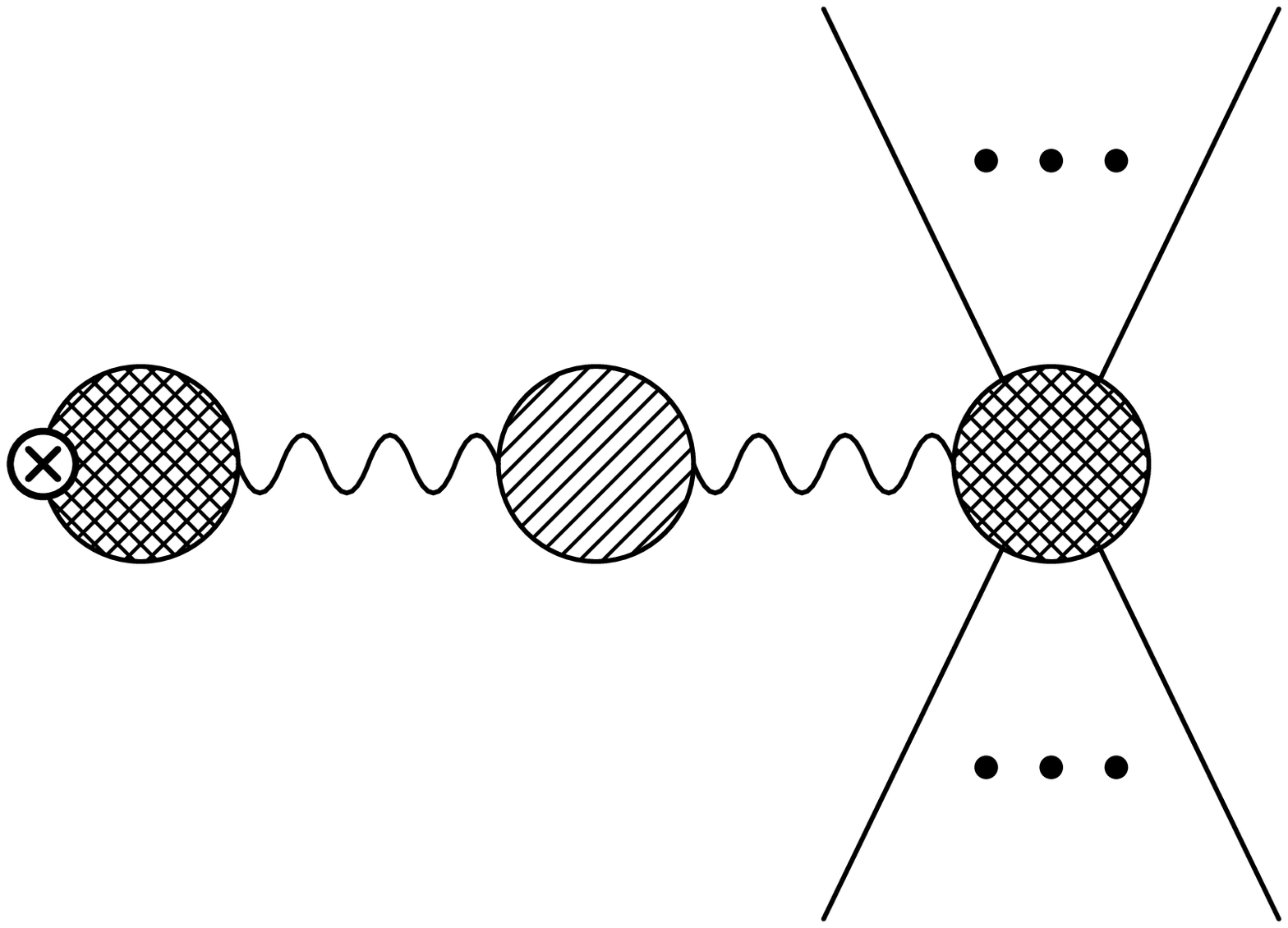}
   & \includegraphics[scale=0.25]{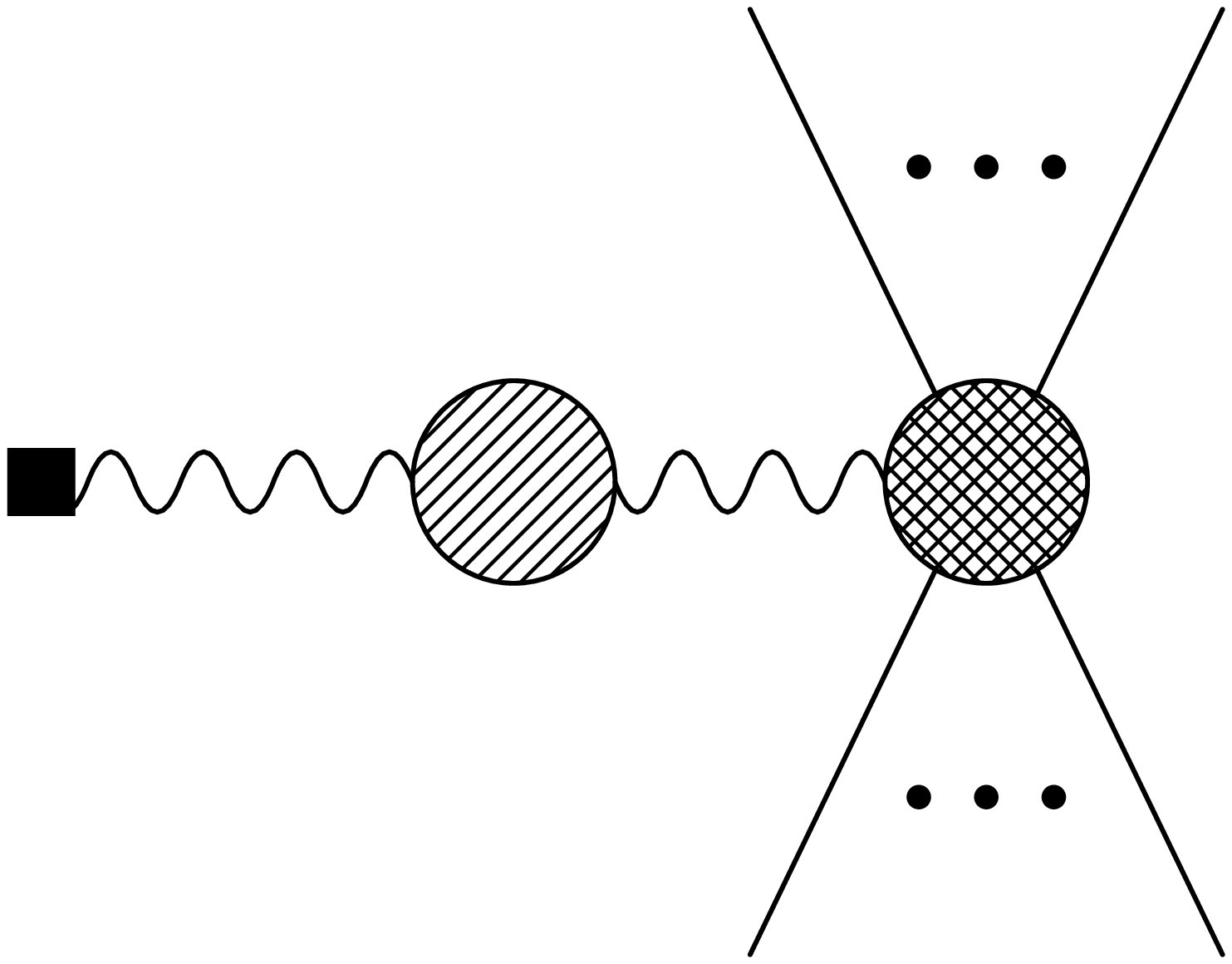}
\\
   (a) & (b) & (c)
\end{tabular}
\caption{(a) Graphs that are one-particle-irreducible in the
   current channel for insertion of a current vertex in a Green
   function or matrix element; the standard nonrenormalization
   argument applies only to these. 
(b) These graphs, one-particle-reducible in the current
  channel, also contribute to matrix elements of the current and to its
  renormalization. The two subgraphs that are cross hatched are
  irreducible in the photon line, while the other subgraph gives the
  full propagator corrections to the photon propagator.
(c) Counterterm to (b).  The filled square corresponds to an
      operator proportional to $\partial_\nu F^{\nu\mu}$.
 \label{fig:penguin.general}
} 
\end{figure*}

However in QED (and therefore in the full Standard Model), the current
does mix with the four-divergence of the electromagnetic
field-strength tensor, $\partial_\nu F^{\nu\mu}$.  The mixing is associated with
``penguin graph'' contributions to matrix elements, Fig.\
\ref{fig:penguin.general}(b).  The conventional proof \cite{sterman}
of nonrenormalization ignores such graphs.  Similar statements apply
in more general theories \footnote{Thus, for example, the validity of
  textbook statements of non-renormalization of currents for quark
  number, like that in \cite[p.\ 430]{peskin.schroeder}, depends on
  whether the statements are taken in pure QCD or in the full Standard
  Model, which contains a massless photon field.}.

In this article, we explain the necessary modifications, to all orders
of perturbation theory, for the electron number current in QED with a
single lepton flavor.  (The results generalize readily to multiple
lepton flavors.)  There is a series of inter-related results:
\begin{itemize}
\item The current needs ultraviolet (UV) renormalization, by the addition of a
  counterterm proportional to $\partial_\nu F^{\nu\mu}$.  The
  coefficient can be set in terms of the coupling and the $Z_3$
  coefficient in the Lagrangian.
\item The renormalization is by addition of an operator that does not
  affect the Ward identities.  Nevertheless, after the application of
  equations of motion to the renormalized current operator in physical
  matrix elements, the renormalization is effectively multiplicative.
\item If \ms-renormalization is used, the current has a nonzero
  anomalous dimension.
\item Renormalization also affects the normalization of the charge
  operator for electron number, unless the photon has a nonzero mass.
  The main part of this result was first found by Luri\'e \cite[p.\ 371,
  Eq.\ ((8(35))]{lurie}.
\item There is a unique finite correction to the counterterm's
  coefficient that both removes the anomalous dimension and produces a
  correct electron number operator that is the integral of an electron
  number density.  In effect we have to subtract contributions
  associated with vacuum polarization.
\item Thus the Noether current for electron number is not the correct
  operator to define electron number.  A notable illustration is in
  multi-flavor QED, where, for example, the total muon number of a
  one-electron state is nonzero, as computed from the standard Noether
  current.
\item The counterterm restores the Gauss's law relation between total
  charge and the flux of the electric field at infinity.  
\item Although the effect on the total number operator, as opposed to
  the local current, only occurs when the photon mass is exactly zero,
  it depends neither on infrared (IR) divergences as such nor on UV
  divergences. 
  It occurs even in a space-time dimension greater than 4, where there
  are no soft divergences in the scattering matrix, and with a
  spatial-lattice cutoff, which removes all UV problems.
\item A simple standard argument involving equal-time canonical
  anticommutators, appears to show that the electron number of states
  is not renormalized, contrary to reality.  We resolve this paradox,
  which was first found by Weeks \cite{weeks} (see appendix).
\end{itemize}

The possibility that problems exist can be motivated by the standard
proof that the electron-number operator is time-independent:
\begin{equation}
\label{eq:charge.conservation}
  \frac{{\rm d}Q}{{\rm d}t} 
  = \int {\rm d}^3 \mathbf{x} \ \frac{\partial j^0}{ \partial t}
  = - \int {\rm d}^3\mathbf{x} \ \bm{\nabla} \cdot \mathbf{j} 
  = - \int_\infty {\rm d}^2 \mathbf{\Sigma} \cdot \mathbf{j} ,
\end{equation}
where the surface term is usually dropped.  But masslessness of the
gauge field allows a nonzero surface term.  As we will
see in detail, such a term does in fact arise, from penguin graphs,
Fig.\ \ref{fig:penguin.general}(b).  Although the left-hand bubble in
Fig.\ \ref{fig:penguin.general}(b) vanishes quadratically as the
external momentum $q$ at the current vertex goes to zero, there is a
pole at $q^2=0$ in the photon propagator.  In addition, the penguin
graphs need nontrivial UV renormalization, with the counterterm
graphs shown in Fig.\ \ref{fig:penguin.general}(c)

Although our results are quite elementary, and should be well-known,
the only textbook account we have found is by Luri\'e \cite[p.\
371]{lurie}. 
Since current operators are auxiliary operators used to analyze a
theory, the nontrivial renormalization of the current does not have
direct effects on the scattering matrix and cross sections.  Weeks's
paradox does symptomize some deep complications in the correct
definition of states in a gauge theory.  

However, many practical calculations use the operator product
expansion and factorization methods.  Then matrix elements of currents
and related operators are 
used, so that incorrect theoretical expectations for anomalous
dimensions will affect predictions.  Beneke and Neubert
\cite{Beneke.Neubert} indeed found closely related effects in their
analysis of $B$ decays, and remarked (without further comment) that
quark number currents have nonzero anomalous dimensions in the
presence of electromagnetic interactions.

\section{General argument}

We use the standard gauge-fixed Lagrangian density
\begin{align}
  \mathcal{L} ={} &
    \bar\psi^{(0)}  \left( i \gamma\cdot\partial 
                            +e_0 \gamma\cdot A^{(0)}
                            -m_0 
                     \right)
         \psi^{(0)}  
\nn &
   -\frac{1}{4} \left( F_{\mu\nu}^{(0)}  \right)^2
   - \frac{1}{2\xi} (\partial\cdot A)^2
\\ ={}&
    Z_2 \bar\psi  \left( i \gamma\cdot\partial 
                            +e\mu^\epsilon \gamma\cdot A
                            -m_0 
                     \right)
         \psi
\nn &
   -\frac{Z_3}{4} \left( F_{\mu\nu}  \right)^2
   - \frac{1}{2\xi} (\partial\cdot A)^2.
\end{align}
Our conventions are that a superscript ${}^{(0)}$ (or subscript in
$e_0$ and $m_0$) denotes bare quantities, dimensional regularization
is in $4-2\epsilon$ dimensions, and the $\gamma$-matrices in
$n$-dimensions are normalized to obey ${\rm Tr}(\gamma_{\mu}
\gamma_{\nu})=4 g_{\mu\nu}$.  Note that the bare and renormalized
couplings obey $e_0=e\mu^\epsilon Z_3^{-1/2}$, with $\mu$ being the
usual unit of mass, and that the gauge fixing term has no counterterm.
We will use \ms-renormalization for the interactions, although this
will not be essential: one formula for the 
current will be in an explicitly renormalization-scheme-independent
form.  The vacuum matrix element of a time-ordered product of
operators is denoted as $\vev{\dots}$.  (The time-ordered product is
actually the covariant $T^*$ product, as naturally arises with Feynman
graphs or from a functional integral solution of the theory.)

\begin{widetext}
The standard Noether current for electron number is $j_{\rm N}^\mu =
\bar\psi^{(0)} \gamma^\mu \psi^{(0)} =Z_2 \bar\psi\gamma^\mu \psi$,
and it obeys a Ward identity:
\begin{align}
\label{eq:Ward.id}
\frac{\partial}{\partial x^\mu}
    \vev{ j_{\rm N}^\mu(x) 
                   \prod_{i=1}^{m}\psi(y_i) 
                   \prod_{j=1}^{m} \bar\psi(z_j) 
                   \prod_{k=1}^{l} A(w_k) 
    }
={}& -\sum_{i'=1}^m \delta^{(4)}(x-y_{i'}) 
      \vev{
                   \prod_{i=1}^{m}\psi(y_i) 
                   \prod_{j=1}^{m} \bar\psi(z_j) 
                   \prod_{k=1}^{l} A(w_k) 
       }
\nn&
+\sum_{j'=1}^m \delta^{(4)}(x-z_{j'}) 
      \vev{
                   \prod_{i=1}^{m}\psi(y_i) 
                   \prod_{j=1}^{m} \bar\psi(z_j) 
                   \prod_{k=1}^{l} A(w_k) 
       } .
\end{align}
\end{widetext}
Because this formula is homogeneous in the fields, it applies equally
whether the fields used in the Green function with the Noether current
are bare or renormalized fields; but the normalization of the Noether
current itself is fixed.  Finiteness of the right-hand side might lead
one to conclude incorrectly that Green functions (and hence matrix
elements) of $j_{\rm N}^\mu$ are finite, so that the current is not
renormalized.  Certainly an extra multiplicative factor is excluded
\cite{sterman,pokorski,collins}, and a closely related argument proves that the
sum of nonpenguin graphs Fig.\ \ref{fig:penguin.general}(a) is
finite: the factor $Z_2$ in $Z_2 \bar\psi\gamma^\mu \psi$ takes care
of divergences in these graphs.

But it is possible to have a counterterm proportional to an operator
whose four-divergence vanishes identically (i.e., without the use of
the equations of motion).  If the equations of motion were needed to
prove the vanishing of the four-divergence of a counterterm, extra
divergent terms would appear on the right-hand side of the Ward
identity.

In QED there is available a possible counterterm operator with the
appropriate dimension and symmetry properties: $\partial_\nu
F^{\nu\mu}$.  The associated divergence is in the left-hand bubble of
Fig.\ \ref{fig:penguin.general}(b), with its vertex for the Noether
current.  It is evidently closely related to vacuum polarization and
hence to the wave function renormalization factor $Z_3$ for the photon
field.  The counterterm operator appears in graphs of the form of
Fig.\ \ref{fig:penguin.general}(c).  Its coefficient can be determined
by the operator equation of motion for the renormalized photon field:
\begin{equation}
\label{eq:e.of.m}
  0 = \frac{\delta S}{\delta A_\mu(x)}
    = e\mu^\epsilon j_{\rm N}^\mu + Z_3 \partial_\nu F^{\nu\mu}
      + \frac{1}{\xi} \partial^\mu \partial \cdot A,
\end{equation}
where $S=\int {\rm d}^{4-2\epsilon}x \, \mathcal{L}$ is the action for
the theory.  Hence, uniqueness of expansions in poles at $\epsilon=0$
shows that the \ms-renormalized current is
\begin{equation}
\label{eq:msbar.current}
j^\mu_{\ms} = Z_2 \bar \psi \gamma^\mu \psi 
              + \frac{ Z_3 - 1 }{ e \mu^{\epsilon} }
                \partial_\nu F^{\nu \mu},
\end{equation}
in terms of which the photon equation of motion has a simple
form in terms of finite operators:
\begin{equation}
\label{eq:e.of.m.ren}
  0 = e\mu^\epsilon j_{\ms}^\mu + \partial_\nu F^{\nu\mu}
      + \frac{1}{\xi} \partial^\mu \partial \cdot A.
\end{equation}

Now the use of the operator equations of motion induces extra delta
function terms when applied in Green functions.  However this is not
the case in matrix elements.  Moreover, the Gupta-Bleuler condition
gives zero physical matrix elements for $\partial\cdot A$.  So we can
eliminate either one of the operators in Eq.\ (\ref{eq:msbar.current})
in favor of the other, plus a term that vanishes in physical matrix
elements:
\begin{align}
  j^\mu_{\ms}
={}& 
  -\frac{ \partial_\nu F^{\nu \mu} }{ e\mu^\epsilon }  
  - \frac{1}{ \xi e\mu^\epsilon } \partial^\mu \partial \cdot A.
\\
={}& 
\frac{1}{Z_3} j_{\rm N}^\mu
  - \frac{ 1-Z_3^{-1} }{ \xi e\mu^\epsilon } \partial^\mu \partial \cdot A.
\end{align}
Thus in physical matrix elements, the renormalized current obeys:
\begin{equation}
\label{eq:msbar.current.physME}
j^\mu_{\ms}
= -\frac{ \partial_\nu F^{\nu \mu} }{ e\mu^\epsilon }
= \frac{1}{Z_3} j_{\rm N}^\mu
\qquad\mbox{(in physical ME)}.
\end{equation}
The formulae with only gauge fields exhibit the finiteness of the
renormalized current, while those with the Noether current exhibit the
nontrivial renormalization of the current

One further useful result is the expression of the current in terms of
bare fields:
\begin{equation}
\label{eq:msbar.current2}
  j^\mu_{\ms}
  =
  \bar \psi^{(0)} \gamma^\mu \psi^{(0)} 
              + \frac{ 1 - Z_3^{-1} }{ e_0 }
                \partial_\nu F^{(0)\, \nu \mu}.
\end{equation}

The multiple formulae for the current lend themselves to different
interpretations.  Within Feynman graph calculations, the distinction
between a counterterm proportional to $\partial_\nu F^{\nu\mu}$ and
one proportional to $j_{\rm N}^\mu$ is absolutely clear and
unambiguous: Thus the left-hand bubble of Fig.\ 
\ref{fig:penguin.general}(b) is renormalized using the $\partial_\nu
F^{\nu\mu}$ operator, not $j_{\rm N}^\mu$.  But after applying the
equations of motion, the distinction is not so clear cut.  Indeed,
using the formula for the renormalized current in terms of the Noether
current will lead us to Weeks' paradox in Sec.\ \ref{sec:paradox}.

\section{One-loop verification}
\label{sec:oneloop}

\begin{figure}
\centering
\includegraphics[scale=0.25]{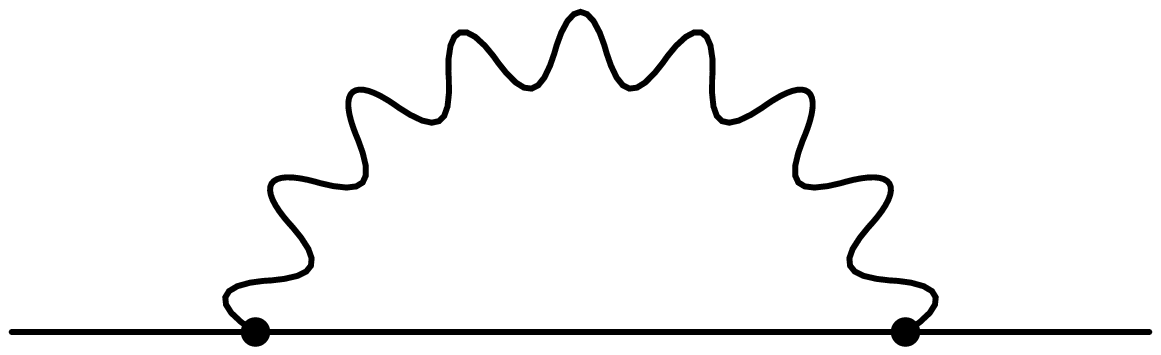}\hspace{0.5cm}
\includegraphics[scale=0.25]{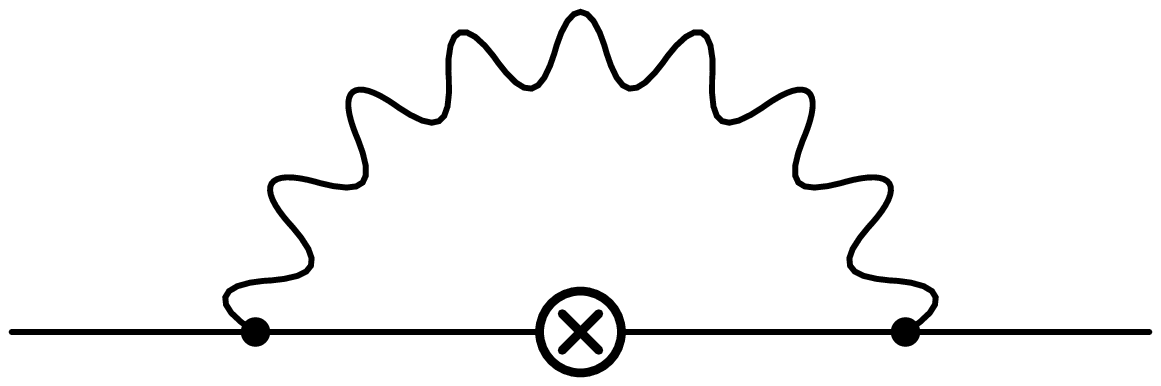}
\caption{One-loop wave-function and vertex corrections. \label{fig:wv}}
\end{figure}

\begin{figure}
\centering
\includegraphics[scale=0.25,angle=90]{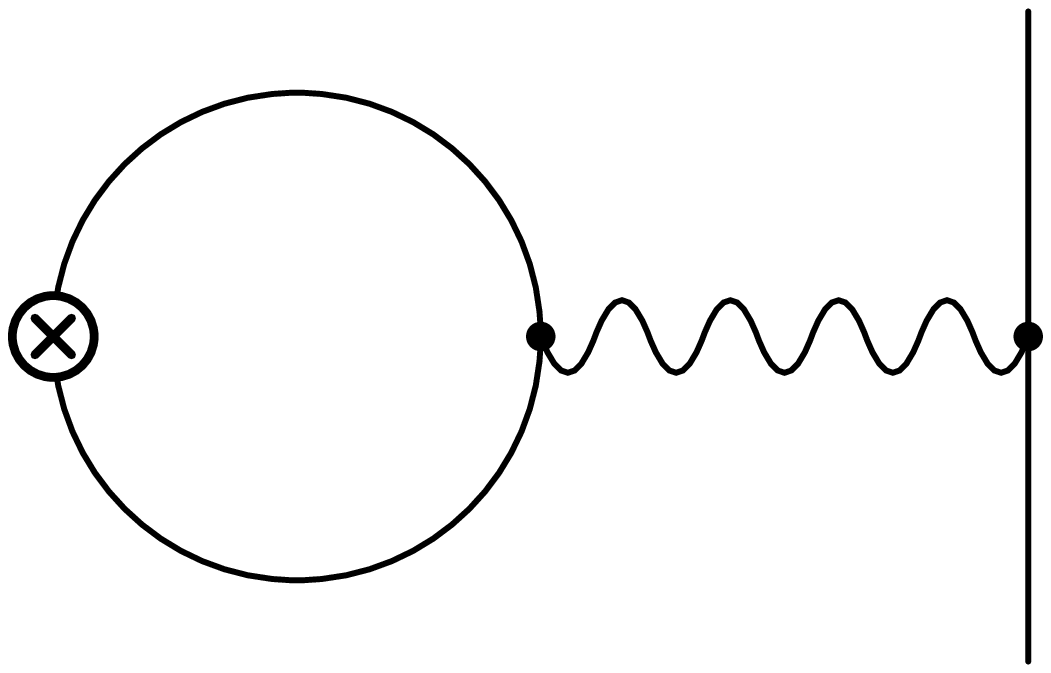}
\caption{One-loop penguin graph contributing to the renormalization of
  the vector current. The cross denotes an insertion of the vector
  current. \label{fig:penguin}} 
\end{figure}

In a one-electron state, the one-loop matrix element of the electron
number current has contributions from wave-function and vertex graphs,
Fig.~\ref{fig:wv}, and from a penguin graph Fig.~\ref{fig:penguin}.
As is well-known, the UV-infinite parts of the wave-function and
vertex graphs cancel.  Moreover, at zero momentum transfer in an
on-shell state, the complete vertex and wave function graphs cancel,
so that they make no contribution, for example, to the expectation
value of the electron number operator.

However, the penguin graph also contributes.  Its 1PI part is
\begin{align}
\label{eq:penguin.1}
   & \frac{e S_\epsilon \mu^{- \epsilon} }{ 2 \pi^2 }
   \left( -q^2 g^{\mu \nu} + q^\mu q^\nu \right) \times
\nn
   &\left\{
      \frac{1}{6\epsilon}
      - \int_0^1 {\rm d}x\ x(1-x) 
            \ln \left[ \frac{m^2-q^2 x(1-x)}{\mu^2} \right]
      + \mathcal{O}(\epsilon)
   \right\},
\end{align}
where $S_\epsilon=(4\pi e^{-\gamma_E})^\epsilon$.  The graph has a
divergence which is canceled by the \ms-counterterm in Eq.\ 
(\ref{eq:msbar.current}) with
\begin{equation}
  Z_3 = 1 - \frac{e^2S_\epsilon}{12\pi^2 \epsilon} 
      + \mathcal{O}(e^4).
\end{equation}
Here we use the usual QCD definition of \ms-renormalization that the
counterterms contain a factor of $S_\epsilon$ for each loop.

\section{Apparent Electron Number Anomaly}

From our calculations, we can see that the penguin graph contribution
to the matrix element of the current apparently changes the value of
electron number, no matter whether we use the Noether current itself
or the \ms-renormalized current.  For the total electron number we
need the limit as $q^2\to0$.  Although the loop vanishes at $q^2=0$,
this is canceled by the pole in the photon propagator.  Thus the
matrix element of $j^\mu_{\ms}$ in a single-electron state at zero
momentum transfer (and $\epsilon=0$) is
\begin{eqnarray}
\langle e|j^\mu_{\ms}|e\rangle
= \left[ 1 + \frac{e^2}{12 \pi^2} 
             \ln\left( \frac{m^2}{\mu^2}\right)
           + \mathcal{O}(e^4)
  \right]
  2 p^\mu
\end{eqnarray}
so that the electron number of an electron is apparently
\begin{eqnarray}
\label{eq:ren.factor}
   1+ \frac{e^2}{12 \pi^2} \ln\left(\frac{m^2}{\mu^2}\right) + \mathcal{O}(e^4).
\end{eqnarray}
This is not even renormalization-group invariant.
Therefore we cannot interpret the current as corresponding to electron
number in the standard way, despite the fact that the current has the
correct Ward identity that corresponds to the standard commutation
relations between the current and the fields.  

The structure of the graphs in Fig.\ \ref{fig:penguin.general} shows
that the factor (\ref{eq:ren.factor}) is universal between different
matrix elements of the current and charge.

An even more dramatic and paradoxical consequence occurs if we add a
second flavor of fermion, which we can call a muon.  The matrix
element of the muon-number current has a contribution from a graph of
the form of Fig.\ \ref{fig:penguin}, with the electron loop replaced
by a muon loop.  Thus the muon number of the electron appears to be
nonzero.

It is also useful to examine the RG properties of the \ms-renormalized
current.  Its anomalous dimension can be related to the RG of the
interaction by use of Eq.\ (\ref{eq:msbar.current2}):
\begin{equation}
\label{eq:current.rge}
 \mu \frac{ {\rm d} }{ {\rm d}\mu } j^\mu_{\ms}
  =
  2\gamma_A \frac{ \partial_\nu F^{(0)\, \nu \mu} }{ e_0 Z_3 }
  = 
  2\gamma_A \frac{ \partial_\nu F^{\nu\mu} }{ e\mu^\epsilon },
\end{equation}
where $\gamma_A$ is the anomalous dimension of the photon field:
\begin{eqnarray}
  \gamma_A(e) 
  = \frac{1}{2 Z_3} \mu \frac{ {\rm d} Z_3 }{ {\rm d} \mu } 
  = \frac{ e^2 S_\epsilon }{ 12 \pi^2 } + \mathcal{O}(e^4), 
\end{eqnarray}
which is related to the QED $\beta$-function by
\begin{eqnarray}
\mu \frac{ {\rm d} e }{ {\rm d} \mu } = - \epsilon e + \beta(e)
 = - \epsilon e + e \gamma_A(e).
\end{eqnarray}

The nonzero anomalous dimension of the current, in
Eq.~(\ref{eq:current.rge}), is consistent with the renormalized
equations of motion, Eq.\ (\ref{eq:e.of.m}).  We see this by use of
the relations between $\beta$ and $\gamma_A$, together with the
anomalous dimension of the gauge-fixing parameter, ${\rm d}\xi/{\rm
  d}\mu = - 2\gamma_A$:
\begin{align}
  0={}&
  \mu \frac{ {\rm d} }{ {\rm d} \mu }
  \left( 
      \partial_\nu F^{\nu\mu} + e \mu^\epsilon j^\mu_{\ms}
      + \frac{1}{\xi} \partial^\mu \partial \cdot A
  \right)
\nn ={}&
  \gamma_A
  \left( 
      \partial_\nu F^{\nu\mu} + e \mu^\epsilon j^\mu_{\ms}
      + \frac{1}{\xi} \partial^\mu \partial \cdot A
  \right).
\end{align}
The right-hand side is proportional to the equation of motion, so that
it vanishes.  Finally, applying Eq.~(\ref{eq:msbar.current.physME}) gives a
renormalization group equation for the current in terms of itself when
physical matrix elements are taken:
\begin{equation}
\mu \frac{ {\rm d} }{ {\rm d} \mu } j^\mu_{\ms} =-2\gamma_A\  j^\mu_{\ms}
\qquad\mbox{(in physical ME)}.
\end{equation}

\section{Redefinition of current}

Since the apparent anomaly arises from penguin-graph contributions
only when the photon is exactly massless, a more satisfactory
definition of the current evidently requires us to remove the penguin
graph contributions.  To preserve the locality of the current, we can
perform an exact removal only at one value of $q$, naturally $q=0$.
We will demonstrate that the correct definition (only valid if the
electron mass $m$ is nonzero) is:
\begin{align}
\label{eq:phys.current.def}
  j^\mu ={}& j^\mu_{\ms} 
          - \frac{ \Pi(0) \,  \partial_\nu F^{\nu\mu} }
                 { e\mu^\epsilon }
\nn
        ={}& j^\mu_{\rm N} 
          + \left[ Z_3 - 1 - \Pi(0) \right]
            \frac{ \partial_\nu F^{\nu\mu} }
                 { e\mu^\epsilon }
\nn
         ={}& \bar\psi^{(0)} \gamma^\mu \psi^{(0)}
          +  \left[ 1 - \frac{ 1 + \Pi(0)}{Z_3} \right]
           \frac{  \partial_\nu F^{(0) \, \nu\mu} }
                 { e_0 }.
\end{align}
Here $\Pi(q^2)$ is the vacuum polarization, defined as usual so that
the renormalized photon propagator is
\begin{equation}
\label{eq:photon.prop}
    \frac{i \left( -g_{\mu\nu} + q_\mu q_\nu/q^2\right) }
         { q^2 \, [1 + \Pi(q^2) ] }
  - \frac{ i q_\mu q_\nu \xi }{ (q^2)^2 }.
\end{equation}
The first form of $j^\mu$ in Eq.\ (\ref{eq:phys.current.def}) shows that it
is a finite operator that obeys the standard Ward identity, Eq.\ 
(\ref{eq:Ward.id}).  In the last form, the factor $[1+\Pi(0)]/Z_3$ is
the inverse of the photon-pole residue in the propagator of the
\emph{bare} photon field.  Thus the last form is written solely in
terms of bare quantities, so that the current is RG invariant.
Moreover we have a formula where a nontrivial correction is manifestly
needed even if there are no UV divergences, as when $n<4$.

The extra photon term in the definition of the current depends on the
dynamics of QED, so we call it the dynamical term in the current.  Its
normalization is only known after the theory is solved, and depends on
specifically quantum mechanical effects.  

In physical matrix elements, the equation of motion for the photon
field gives
\begin{align}
\label{eq:phys.current.physME}
j^\mu ={}& -\frac{ [1+\Pi(0)] \ \partial_\nu F^{\nu \mu} }
              { e \mu^\epsilon }
\nn
  ={}&
    [1+\Pi(0)] \ j^\mu_{\ms}
\nn
  ={}&
    \frac{ 1+\Pi(0) }{ Z_3 } \ j^\mu_{\rm N}
\nn
 ={}& -\frac{ \partial_\nu F^{\nu \mu}_{\rm phys} }
              { e_{\rm phys} }
\qquad\mbox{(all in physical ME)}.
\end{align}
In the last line, we use what we term a physically normalized
field, 
\begin{eqnarray}
F^{\nu \mu}_{\rm phys} &=& [1+\Pi(0)]^{1/2} F^{\nu \mu},
\end{eqnarray}
whose photon pole has unit residue.  The corresponding coupling is
\begin{eqnarray}
e_{\rm phys} &=& [1+\Pi(0)]^{-1/2} e \mu^\epsilon,
\end{eqnarray}
which has the conventional value, i.e.,
$1.6022\dots\times10^{-19}\,{\rm C}$, rather than the \ms-renormalized
coupling $e$ that is appropriate in certain high-energy
calculations. Since, with \ms-renormalization, 
\begin{equation}
\label{eq:Pi}
  \Pi(0) = 
   - \frac{ e^2 }{ 12 \pi^2 }
      \ln \frac{m^2}{\mu^2} 
   + \mathcal{O}(e^4),
\end{equation}
we verify that the redefinition Eq.~(\ref{eq:phys.current.def})  does
indeed remove the one-loop anomaly in the electron number. 

To see that the current $j^\mu$ defined in Eq.\ 
(\ref{eq:phys.current.def}) is the uniquely correct current, we
express an arbitrary Green function or matrix element of $j^\mu$ in
terms of $\jir^\mu$, which is the 1-photon-irreducible part in the
current channel.  The reducible graphs are those shown in Fig.\ 
\ref{fig:penguin.general}(b) and (c), so that the total Green function
is
\begin{equation}
\label{eq:jtot.jir}
  \jtot^\mu
=
  \jir^\mu  \frac{ 1 + \Pi(0) }{ 1 + \Pi(q^2) }
  + \jir^\nu \frac{q_\nu q^\mu}{q^2}
    \frac{ \Pi(q^2) - \Pi(0) }{ 1 + \Pi(q^2) }.
\end{equation}
Multiplying by $q_\mu$ gives just the irreducible contribution
$q_\mu\jir^\mu$, which verifies that the Ward identities are
unaffected by the extra terms used to define the physical current.
But when we take a physical matrix element, current conservation shows
that only the first term in Eq.\ (\ref{eq:jtot.jir}) survives, so that
the matrix element differs from the irreducible contribution by a
factor $[1 + \Pi(0)] / [1 + \Pi(q^2)]$, which is unity when $q^2=0$.
The standard arguments about nonrenormalization of the current
actually apply only to the 1-photon-irreducible part, Fig.\ 
\ref{fig:penguin.general}(a), which is finite and unaffected by the
addition of the dynamical photon term to the current.

To see that the extra dynamical term in the current is due to bad
behavior of the current operator at spatial infinity, we can use a
nonzero photon mass.  Then the factor between the total and
irreducible parts of a matrix element of the current becomes
\begin{align}
  \frac{ q^2-m_\gamma^2 + q^2 \Pi(0) }
       { q^2-m_\gamma^2 + q^2 \Pi(q^2) }
 = 
  1+ 
  \frac{ q^2 \, [\Pi(0) - \Pi(q^2)] }
       { q^2-m_\gamma^2 + q^2 \Pi(q^2) },
\end{align}
where the last term, proportional to $\Pi(0) - \Pi(q^2)$, is the
contribution of the penguin graphs and of the dynamical term.  Because
of the photon mass in the denominator, both of these contributions
vanish at $q=0$.  

Finally, we observe that the formula (\ref{eq:phys.current.physME}) for the
physical current in terms of the electromagnetic field strength shows
that the time-component of the current is 
\begin{equation}
   j^0 = \frac{ \bm{\nabla} \cdot \mathbf{E} }{ -  e_{\rm phys} }.
\end{equation}
This is the divergence of the electric field operator,
with its \emph{standard} normalization, in units of the (negative)
charge of the
electron.  Integrating over all space shows that the electron number
with this definition equals the value given by Gauss's law with
the standard normalization.  This naturally matches with the
classical limit of electromagnetism, for macroscopic phenomena.

We have defined a physical charge and current, but the necessity of
nontrivial renormalization depends on whether the photon is massless
or not and on whether the current or charge is considered:
\begin{itemize}
\item In the $3+1$ dimensional theory, the Noether current always
  needs UV renormalization, independently of $m_\gamma$.
\item The counterterm never affects the Ward identities.
\item When the photon mass is nonzero, the counterterm integrates to
  zero, so that the charge does not then actually need
  renormalization, and the integral of the Noether charge density
  gives the correct charge.
\item But when the photon mass is zero:
  \begin{itemize}
  \item Integrating the counterterm over all space gives a nonzero
    result, so that the charge needs UV renormalization.
  \item A particular finite part in the counterterm is needed to
    obtain the correct charge.
  \item Even when the theory is regulated in the UV, the finite
    renormalization of the current and charge are still needed, if the
    the charge is to be correct.
  \end{itemize}
\item However, even with a nonzero photon mass, the definition of the
  physical current (\ref{eq:phys.current.def}) is valid and gives the
  correct charge.  (The counterterm integrates to zero.)  This
  definition has the practical advantages that there are no problems
  in taking the limits of zero photon mass, integration over all
  space, and removing a UV regulator, and that changing the order of
  limits is safe.
\end{itemize}

Evidently the need for a nontrivial renormalization of the total
electron number (as opposed to the current) only arises when the
photon is massless.  However, it is to be emphasized that this is not
due to actual infra-red divergences as normally understood.  This can
be seen by considering the theory in 5 space-time dimensions (i.e., 4
space dimensions), with a spatial lattice.  The higher space-time
dimension is sufficient to remove all the usual soft divergences in
the scattering matrix.  The use of an integer dimension removes all
possible artifacts associated with dimensional regularization, and the
use of a spatial lattice gives us a conventional quantum mechanical
theory (in real time) without UV divergences.  All the considerations
leading to a physical current that is not equal to the Noether current
still apply.

It is sometimes claimed that the actual definition of QED requires us
first to regulate the theory in the IR, say with a nonzero photon
mass, and then to take the limit of zero photon mass.  If this were
so, we could avoid the need to redefine the charge, since in the
regulated theory the Noether charge would give the correct answer.
However, as a field theory, QED exists if the photon mass is kept at
zero at all stages; this is evidenced by the fact that the Green
functions of the theory exist directly at $m_\gamma=0$.  The use of an IR
regulator (and inclusive cross sections that allow undetected soft
photon emission) is only necessary if one uses the conventional LSZ
formalism to compute cross sections.  In any case, when the space-time
dimension is above 4 there are no soft photon divergences in cross
sections, but nevertheless a redefinition of the charge is still
needed, as we have just seen.

Our arguments generalize to more complicated theories.  For example,
the approach applies to QED with more than one flavor of lepton, with
suitable changes in the numbers of flavors in the vacuum polarization
graphs.  The approach applies not only to the electromagnetic current
itself, but also to the individual conserved flavor-number currents. 

We propose an interpretation of the nontrivial renormalization of the
current as a universal effect of attempts to measure the current.  The
quantum mechanical current creates polarization of the vacuum.
Although this effect is proportional to $1/r^2$ at a distance $r$ from
a source, it has to be integrated over a sphere of surface area
proportional to $r^2$, so that the effect is nonvanishing as
$r\to\infty$.

\begin{acknowledgments}
  This work was supported in part by DOE grants DE-FG02-90ER-40577,
  DE-FG03-97ER40546 and DE-FG03-92ER40701.
  We thank S. Adler, M. Neubert, J. Rabin, M. Srednicki, and
  M. Voloshin for comments on the first version of this paper. 
\end{acknowledgments}


\appendix

\section{Weeks' paradox}
\label{sec:paradox}

\subsection{Formulation}

Conventional expectations for the interpretation of a charge operator,
and, in particular, for its value in a state of definite electron
number arise quite generally from the commutation relations between
the Noether current and bare elementary fields.  These commutation
relations are entirely unaffected by UV renormalization and are
fundamental to the definition of a quantum field theory.  Inspired by
the paper of Weeks \cite{weeks}, we now present a paradox.  

The paradox and its resolution turn on the issue of the correct way to
implement concretely the concept of the electron number of a state.
There are several possible ways of implementing the concept.  In
scattering theory, we can use in- or out-states as basis states; the
labeling of these states in the usual way gives unambiguous values for
electron number.  But we could also define electron number in terms of
eigenvalues of a suitable operator for conserved charge.  Yet another
possible definition would use the expectation value of the charge in a
state.  In the absence gauge fields, the different definitions can be
proved to agree.  But, as we will see, in QED an attempt to make such
a proof runs into problems.  The problems are related to, but distinct
from, the issues treated in the body of this paper.

The paradox is obtained as follows:
\begin{enumerate}
\item Let $|0\rangle$ be the (true) vacuum state and let $|1\rangle$ be any state
  of electron number unity.  While we may use an on-shell one-electron
  state of definite momentum for $|1\rangle$, it is safer to use a
  normalizable wave-packet state.  Here we chose the states to be
  normal in- or out-states, with electron number being defined with
  respect to the obvious labeling of the states in terms of particle
  content.
  
\item We assume, as is natural, that $\langle1|\bar\psi|0\rangle \neq 0$, and that $|0\rangle$
  and $|1\rangle$ are eigenstates of the Noether charge $Q_{\rm N}$.  Of
  course, we have shown in the body of this paper that there is an
  unexpectedly nontrivial renormalization of the charge, which leads
  to the expectation that the eigenvalue disagrees with naive
  expectations.  But this finding by itself does not impinge on the
  presumption that we have an eigenstate.
  
\item We (and Luri\'e \cite{lurie}) have calculated that the expectation
  value value of $Q_{\rm N}$ in the state $|1\rangle$ is $\hat{Z}_3$ .  Here
  $\hat{Z}_3$ is the renormalization factor for the photon field in an
  on-shell field.  It equals $Z_3/[1+\Pi(0)]$ in our earlier notation.
  An explicit one-loop calculation shows that $\hat{Z}_3 \not = 1$.
  Of course, on an eigenstate the expectation value of an operator
  equals the eigenvalue.
  
\item From the canonical equal-time anticommutation relations for the
  fermion field we find
  \begin{align}
  \label{eq:weeks}
  \langle1| \bar\psi(t,\3x) |0\rangle = {}&
  \langle1|\left[Q_{\rm N},\bar \psi(t,\3x)\right]|0\rangle
\nn
  ={} &
     \langle1| Q_{\rm N} \bar\psi |0\rangle - \langle1| \bar\psi\, Q_{\rm N} |0\rangle
\nn
  ={}& \left( \hat{Z}_3 -0\right) \langle1|\bar\psi(t,\3x)|0\rangle
  \end{align}
\end{enumerate}
Since $\hat{Z}_3\neq1$, there is a
contradiction, i.e., there is at least one mistake in the calculation
or the assumptions.

Weeks' \cite{weeks} derivation of the paradox was in Coulomb gauge,
but the main ideas are the same. The basic conflict is between the
canonical commutation relation $\left[Q_{\rm N},\bar \psi\right]=1\cdot \bar
\psi$, and the value $\hat{Z}_3\not=1$ for $Q_{\rm N}$ in the state
$|1\rangle$. 

\subsection{Resolution}

The resolution of the paradox is that physical states are not exactly
eigenstates of the Noether charge operator $Q_{\rm N}$, contrary to
the (unproved) assumption made at step 2 of the derivation of the
paradox.  For example, as we will see, when the charge density
operator $j_{\rm N}^0$ is applied to the vacuum and then integrated
over position, there is a certain contribution that vanishes in the
limit of infinite volume (or equivalently when the momentum transfer
$\3q$ goes to zero).  However, matrix elements of a field operator
with this part of the state have a divergence in the same limit.
Moreover this anomalous contribution is associated with unphysical
parts of the theory: the unphysical states in Feynman gauge, or the
instantaneous Coulomb potential in Coulomb gauge.  Thus it is not
easily visible when one restricts attention to physical matrix
elements; in particular, the anomalous contribution does not affect
the calculation of the expectation value of the charge in physical
state.  In addition, in Coulomb gauge the equal-time commutators and
anticommutators are modified by interactions.

We will illustrate these issues by calculations of the commutators in
two ways.  One is the Bjorken-Johnson-Low (BJL) method \cite{BJL}, which is
applied to time-ordered products of the current and fields, while the
second method is a direct calculation with the intermediate states
used in the commutator --- see Eq.\ (\ref{eq:comm.ME}) below.

\subsubsection{Resolution in Feynman gauge}

Consider first the commutator in covariant gauge between the Noether
current and the electron field.
\begin{align}
\label{eq:comm.ME}
     \langle1| [j_{\rm N}^0(t,\3y), \bar\psi(t,\3x) ] |0\rangle
  ={} &
     \sum_X \langle1| j_{\rm N}^0 |X\rangle \langle X| \bar\psi |0\rangle
\nn &
     - \sum_Y \langle1| \bar\psi |Y\rangle \langle Y| j_{\rm N}^0 |0\rangle.
\end{align}
The sums over intermediate states are over all states, not just over
representatives of the physical subspace.  

The BJL method gives this matrix element of the commutator from 
the Fourier transform of the corresponding matrix element of the
time-ordered product, by taking the limit as $q^0\to\infty$ of $iq^0$ times
the matrix element.  Here $q^\mu$ is the external momentum flowing into
the vertex for $j_{\rm N}^0$.  Thus
\begin{align}
\label{eq:comm.BJL}
     \langle1| [j_{\rm N}^0(t,\3y), \bar\psi(t,\3x) ] |0\rangle
  ={} &
\nn& \hspace*{-3cm}
     \lim_{q^0\to\infty} iq^0
      \int {\rm d}t \, e^{iq^0t}
      \langle1| T\, j_{\rm N}^0(t,\3y), \bar\psi(t,\3x) |0\rangle .
\end{align}

\begin{figure}
  \centering
  \includegraphics[scale=0.25]{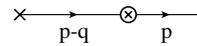}
  \caption{Matrix element on the right of Eq.\ (\ref{eq:comm.BJL})
    used in lowest-order BJL computation of the commutator. The
    circled cross is the vertex for the operator $j_{\rm N}^0$ and the
    simple cross is the vertex for $\bar\psi$.  The external line of
    momentum $p$ is for the state $|1\rangle$. }
  \label{fig:BJL.LO}
\end{figure}

It is readily verified that the lowest-order calculation from the
graph of Fig.\ \ref{fig:BJL.LO} gives the expected commutator:
\begin{equation}
 \lim_{q^0\to\infty} iq^0
  \bar{u} \gamma^0 \frac{ i (\slashed{p} - \slashed{q} + m) }{(p-q)^2-m^2}
 = \bar{u},
\end{equation}
where $u$ is the Dirac wave function for the on-shell one-electron
state $|1\rangle$, of momentum $p$.

\begin{figure}
  \centering
  \includegraphics[scale=0.25]{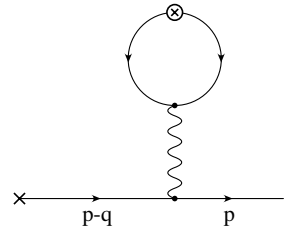}
  \caption{Graph for the one-loop vacuum-polarization contribution to
    the BJL computation.}
  \label{fig:BJL.penguin}
\end{figure}

The one-loop penguin graph contribution to Eq.\ (\ref{eq:comm.BJL}) is
from the graph of Fig.\ \ref{fig:BJL.penguin}.  Any correction factor
to the LO result is from the large $q^0$ behavior of the vacuum
polarization graph times the photon propagator:
\begin{align}
  \frac{ -i g^\mu_\nu }{ q^2 }
  i (-g^{\nu0} q^2 + q^\nu q^0) \Pi(q^2)
 = \left( -g^{\mu0} + \frac{q^\mu q^0}{q^2} \right) \Pi(q^2).
\end{align}
The $\mu=0$ case is
\begin{equation}
  \frac{\3q^2}{q^2} \Pi(q^2),
\end{equation}
which is suppressed at large $q^0$ by two powers of $q^0$.  The
spatial part is
\begin{equation}
  \frac{\3q q^0}{q^2} \Pi(q^2),
\end{equation}
again suppressed, but by one power of $q^0$.

So the canonical commutator is unchanged.  But notice how the $q^\mu q^0$
part of the vacuum polarization is essential to the suppression.  This
contrasts with the fact that this term does not contribute to physical
matrix elements of the current.  We may therefore anticipate that
contributions from unphysical intermediate states in Eq.\
(\ref{eq:comm.ME}) are essential to get the correct result.

\begin{figure}
  \centering
\includegraphics[scale=0.25]{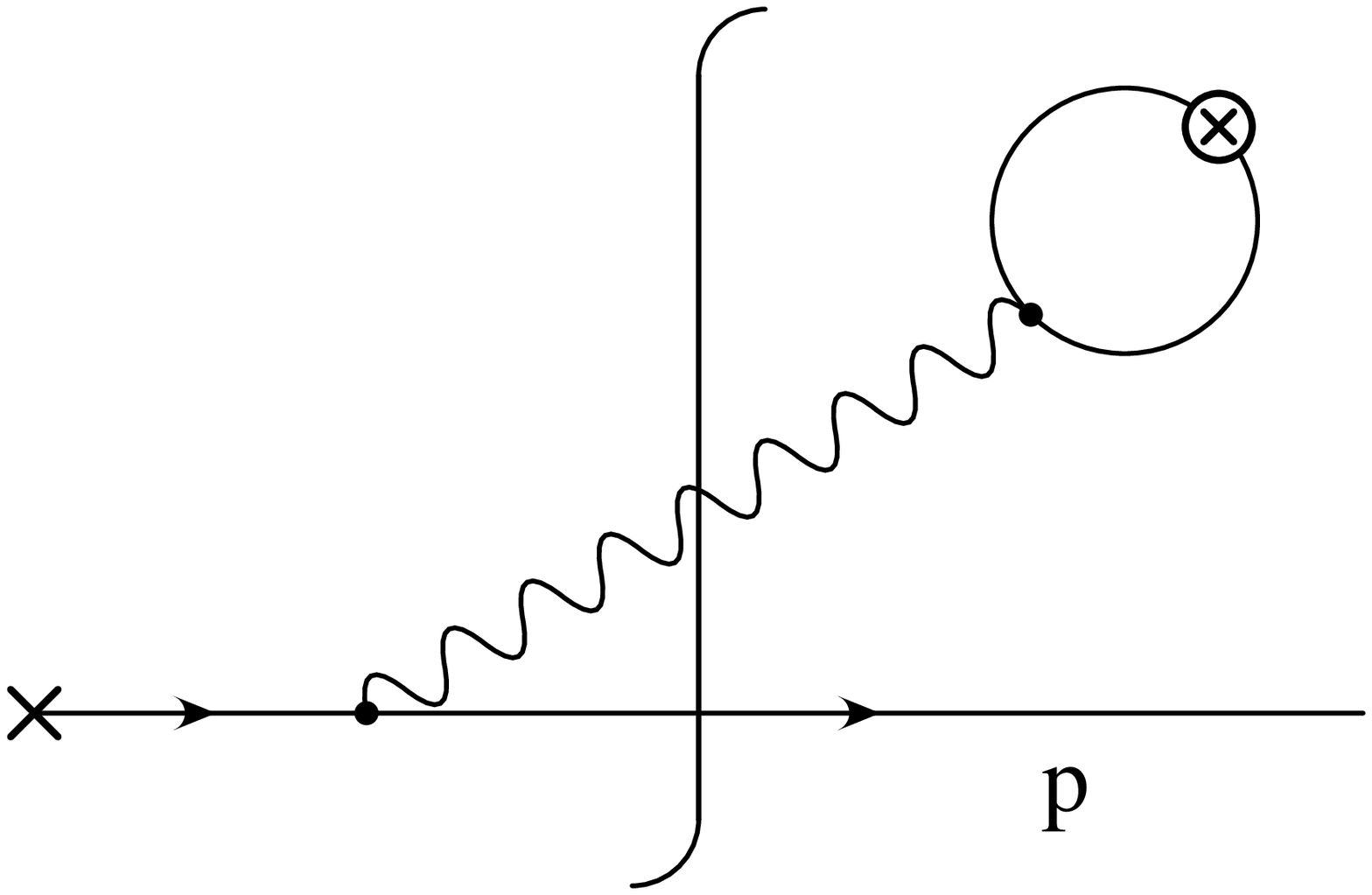}

\vspace{0.5cm}

\includegraphics[scale=0.25]{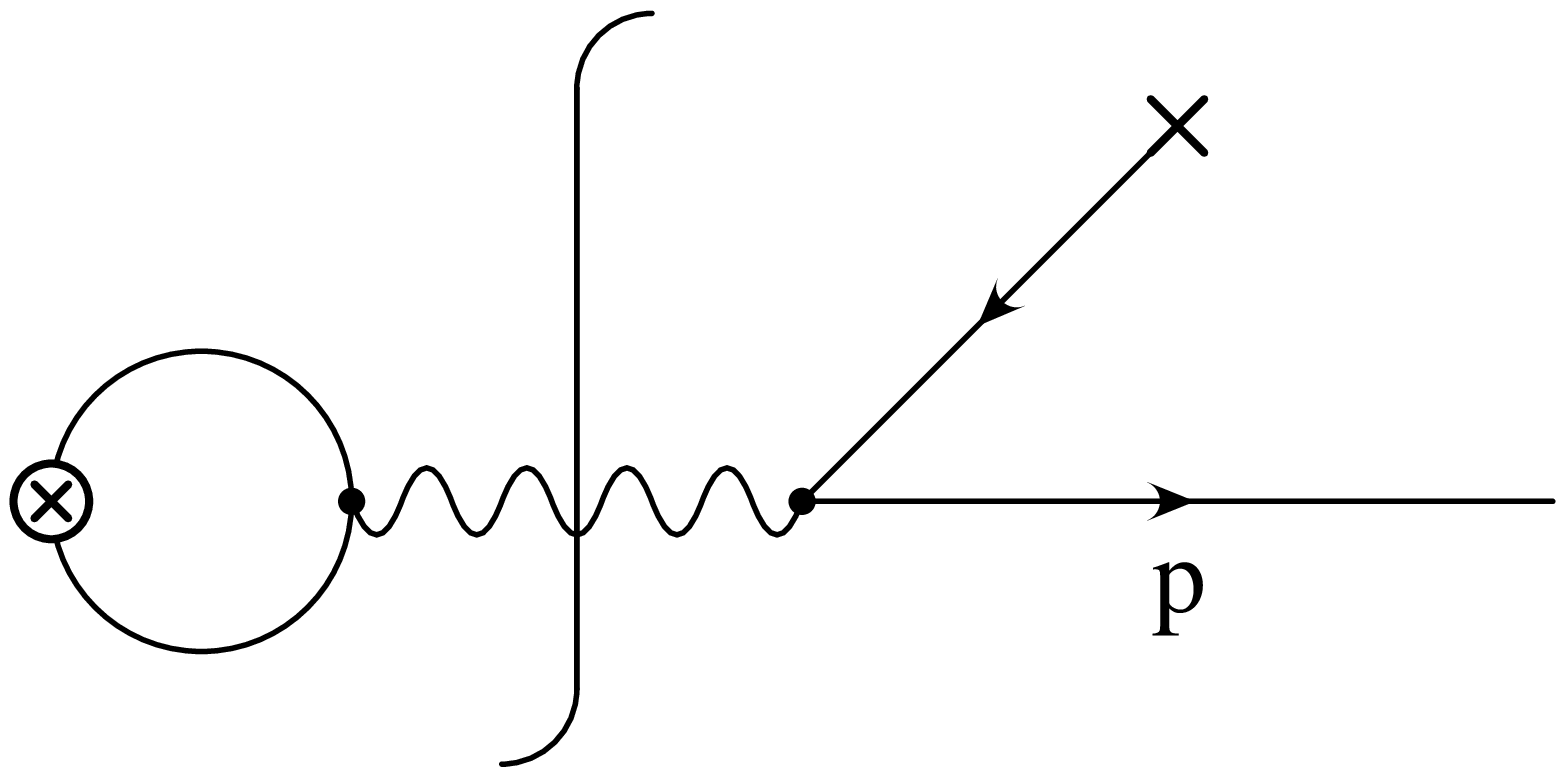}
  \caption{Graphs for the one-loop vacuum-polarization contribution to
    the BJL computation, with cuts on the photon line.}
  \label{fig:BJL.photon.cut}
\end{figure}

\begin{figure}
  \centering
  \includegraphics[scale=0.25]{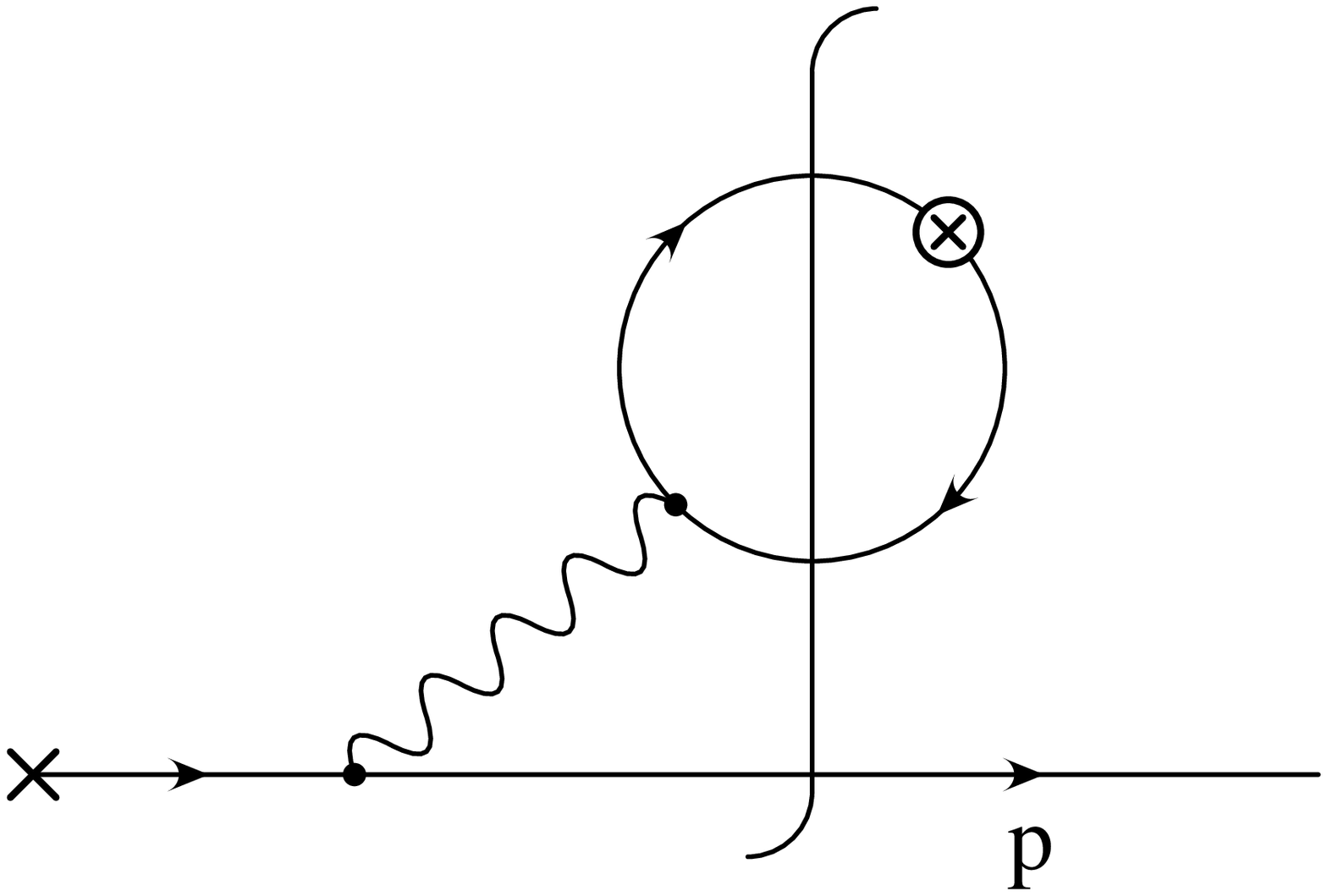}

\vspace{0.5cm}

  \includegraphics[scale=0.25]{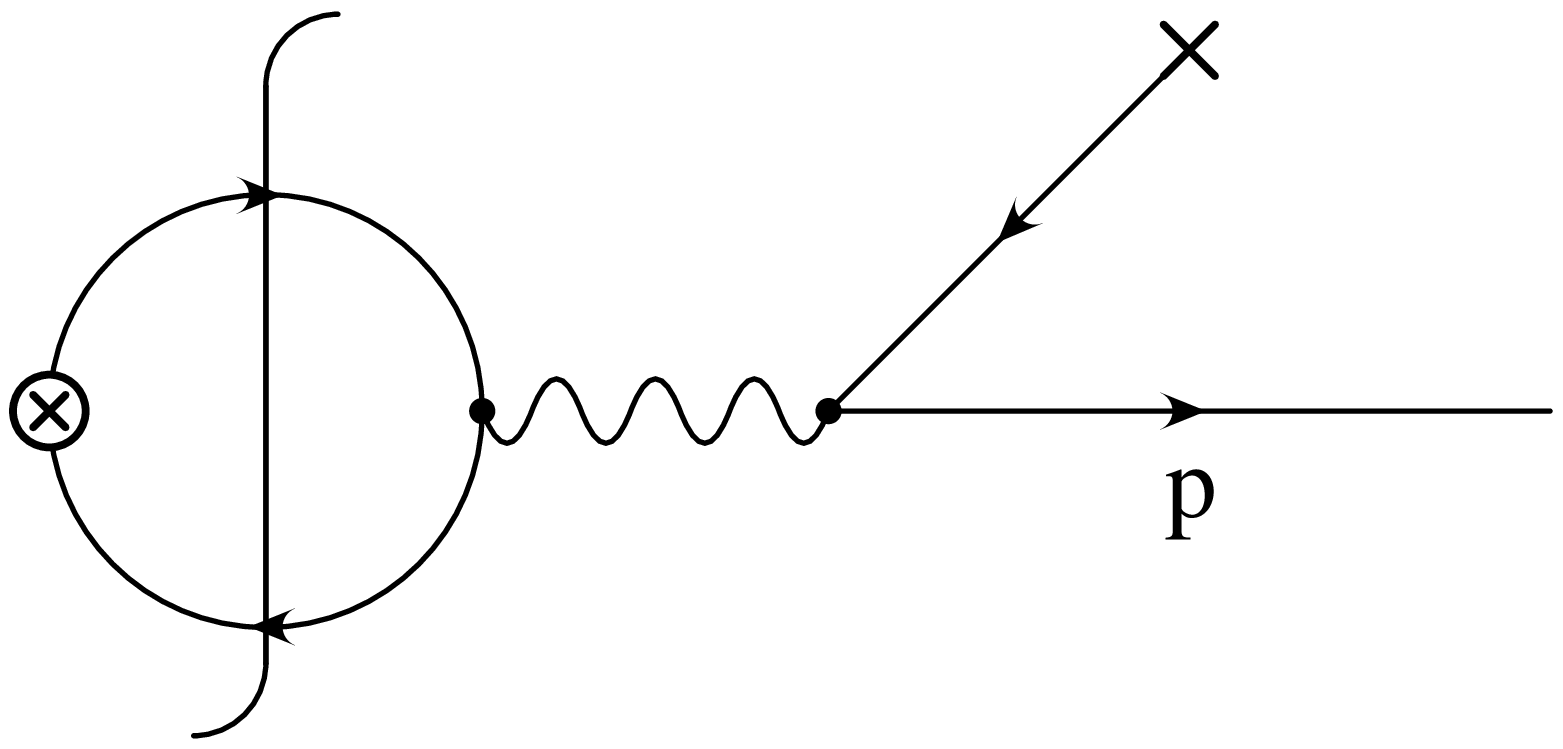}
  \caption{Graphs for the one-loop vacuum-polarization contribution to
    the BJL computation, with cuts on an electron-positron pair.}
  \label{fig:BJL.epem.cut}
\end{figure}

This can be seen explicitly from the graphs of Fig.\
\ref{fig:BJL.photon.cut}, where the cuts correspond to the cases that
$|X\rangle$ is an out-state of one electron and one photon, and that
$|Y\rangle$ is a state of one photon.  The vacuum polarization factor is
\begin{equation}
\label{eq:vac.pol}
  i (-g^{\mu0} q^2 + q^\mu q^0) \Pi(q^2)
=  i q^\mu |\3q|  \Pi(0),
\end{equation}
given that the photon is on-shell, $q^2=0$.  This means that in the
first graph, the sole contribution is from a photon with a scalar
polarization, i.e., a polarization vector proportional to its
lightlike momentum.  This is a state of zero norm: the current applied
to the vacuum has not created a genuine photon.  Moreover, in the
relevant part of (\ref{eq:comm.ME}), the vacuum polarization factor
(\ref{eq:vac.pol}) vanishes quadratically when $\3q\to0$, which is
completely compatible with the natural expectation that the vacuum has
zero charge, so that it is annihilated by the operator $Q_{\rm N}$.

In the commutator, this factor contributes to the matrix element
$\langle Y| j_{\rm N}^0 |0\rangle$.  However it is multiplied by 
$\langle1| \bar\psi |Y\rangle$, and then we get a nonzero result from the
following dependence on $\3q$:
\begin{itemize}
\item The already-derived factor of $\3q^2$.
\item $1/|\3q|$ in the photon relativistic phase space.
\item The IR singularity $\propto1/|\3q|$ in the fermion propagator in
  $\langle1| \bar\psi |Y\rangle$.
\end{itemize}
The nonzero result at $\3q=0$ is independent of the space-time
dimension.  Thus it does not depend on the existence of IR divergences
that violate the standard principles for asymptotic scattering states,
but only on the masslessness of the photon.

To summarize: Although the charge operator applied to the vacuum, for
example, does give zero, this is compensated by an infinity in the
matrix element with the field operator.  Moreover the offending
intermediate state is of zero norm, before taking the $\3q\to0$ limit,
so the state is equivalent to the zero state.  Similar remarks apply
to the one-electron-one-photon intermediate state from the other
diagram in Fig.\ \ref{fig:BJL.photon.cut}.  

A related issue is about the value of the commutator of the
physical charge.  In physical matrix elements, the two charges differ
by a factor $\hat{Z}_3$, and it is tempting to say that the commutators
also differ by this factor.  But consider the expression of the
physical current in terms of the Noether current and $\partial\cdot
A$.  This is obtained by application of the equations of motion
Eq.~(\ref{eq:e.of.m.ren}) to the physical current
Eq.~(\ref{eq:phys.current.def}),  but without the physical state
condition that we used in Eq.~(\ref{eq:phys.current.physME}): 
\begin{align}
j^\mu ={}& 
  \frac{ 1 }{ \hat{Z}_3 } \ j^\mu_{\rm N}
  + \frac{ \hat{Z}_3^{-1} -1 }{ \xi e\mu^\epsilon }
    \partial^\mu \partial \cdot A.
\end{align}
The physical electron number operator is defined as an integral over
$j^0$, so that
\begin{align}
Q_{\rm phys} ={}& 
  \frac{ 1 }{ \hat{Z}_3 } \ Q_{\rm N}
  + \frac{ \hat{Z}_3^{-1} -1 }{ \xi e\mu^\epsilon }
    \int {\rm d}^3\3x \ \partial^0\left( \partial \cdot A\right).
\end{align}
In physical matrix elements we can ignore the second term, to get the
usual ratio of $\hat{Z}_3$ between the physical and Noether charges.
But in commutators with elementary fields, we must include the second
term.  Since this second term involves a \emph{double} time-derivative
of the gauge field, the equations of motion must be applied to write
it in terms of elementary fields and their canonical momenta, if we
are to use canonical commutation and anticommutation
relations.  This gives extra terms, so the commutators of $Q_{\rm
  phys}$ and $Q_{\rm N}$ with elementary fields are not simply related
by $\hat{Z}_3$. 

The appropriate formula to use is, in fact, either of the
two last lines of Eq.\ (\ref{eq:phys.current.def}).  For the $\mu=0$
component we have 
\begin{eqnarray}
Q_{\rm  phys} &=& Q_{\rm N} + \left[1-\hat{Z}_3^{-1}\right] \int {\rm d}^3\3x
    \frac{  \partial_\nu F^{(0) \, \nu 0}}
                 { e_0 } \nn
                 &=& Q_{\rm N}
                 - \frac{1-\hat{Z}_3^{-1}}{e_0}
     \int {\rm d}^3\3x \left[  \3\nabla^2 A^0 
      + \frac{\partial}{\partial t} \3\nabla\cdot\3 A^{(0)}\right], \nn
     \label{eq:div.E}
\end{eqnarray}
i.e., the Noether charge density, with unit normalization,
plus a term that involves at most a first-order time derivative, so
that it commutes at
equal time with the fermion fields, to give $\left[Q_{\rm phys},\bar
  \psi\right]=\left[Q_{\rm N},\bar \psi\right]=\bar\psi$. Thus, despite the
relative factor $\hat{Z}_3$ between the physical and Noether charges
as projected onto physical states, the commutators 
with the fermion do not have this factor.  

\subsubsection{Resolution in Coulomb gauge}

In the Coulomb gauge version of the BJL calculation from the graph of
Fig.\ \ref{fig:BJL.penguin}, the vacuum polarization graph times the
photon propagator is now
\begin{equation}
  \frac{ i (-g^\mu_\nu + q^\mu v_\nu q^0/\3q^2) }{ q^2 }
  i (-g^{\nu0} q^2 + q^\nu q^0) \Pi(q^2),
\end{equation}
where $v^\mu=(1,\30)$ is the unit vector defining the rest frame for the
Coulomb gauge, and we have dropped those terms in the propagator that
are exactly zero because the vacuum polarization is transverse.  It is
easily checked that this formula equals $-g^{\mu0}\Pi(q^2)$.  
As $q^0\to \infty$,
this is missing the power suppression that we have in covariant gauge.
If the theory is regulated in the UV, e.g., by using $\epsilon>0$, then the
vacuum polarization does go to zero, so that the loop correction to
the standard commutator vanishes.  But after the regulator is removed,
the vacuum polarization grows logarithmically.  

Evidently there is an anomaly in the commutator of the Noether charge
density with the field $\bar\psi$.  Because constrained quantization is
used, it is not automatically true that standard canonical commutation
and anticommutation relations can be maintained.  But this result
depends on the space-time dimension.

Interesting dimension-independent pathologies also arise in the
calculation of the commutator with intermediate states, but from
electron-positron terms instead of photon terms.  In the photon term,
we have the same vacuum polarization factor Eq.\ (\ref{eq:vac.pol}) as
before, but now $q^\mu$ times the numerator on the on-shell photon line
gives exactly zero.

Instead there are extra contributions from intermediate states with
electron-positron pairs, Fig.\ \ref{fig:BJL.epem.cut}.  The cut vacuum
polarization graph has the transverse form
\begin{equation}
   (-g^{\nu0} q^2 + q^\nu q^0) \Pi_c(q^2)  
   = \left( \3q^2, \3q\, q^0 \right),
\end{equation}
where $q^2$ is no longer zero, but is bounded below by $4m^2$.  The
spatial part of this vector gives zero when multiplied into the photon
propagator.  However, the time component multiplies a factor $i/\3q^2$
for the instantaneous Coulomb potential, so that we get a nonzero
limit at $\3q=0$.  Of course, the energy $q^0$ is nonzero, but this
is allowed, since the charge operator is computed from the charge
density at a fixed time, so that in momentum space there is an
integral over all $q^0$.

The result of a nonzero limit at $\3q=0$ is dimension independent, and
again indicates that states of definite particle number are not actual
eigenstates of the Noether charge.  But the source of the anomaly has
changed.  In Feynman gauge the anomalous contribution was associated
with unphysical states.  But in Coulomb gauge, it is associated with
the unphysically instantaneous potential in the photon propagator.


\end{document}